\documentclass{aastex}
\usepackage{spr-astr-addons}
\usepackage{url}

\RequirePackage{color}

\begin{document}

\title{Revisiting the classical electron model in general relativity}

\shorttitle{Revisiting the classical electron model}
\shortauthors{Rahaman et al.}

\author{Farook Rahaman\altaffilmark{1}}
\and
\author{Mubasher Jamil\altaffilmark{2}}
\and
\author{Kaushik Chakraborty\altaffilmark{3}}

\altaffiltext{1}{Department of Mathematics, Jadavpur University,
Kolkata, India. Email: farook$\_$rahaman@yahoo.com}
\altaffiltext{2}{Center for Advanced Mathematics and Physics,
National University of Sciences and Technology, Islamabad, Pakistan.
Email: mjamil@camp.nust.edu.pk} \altaffiltext{3}{Department of
Physics, Government Training College, Hooghly - 712103, West Bengal,
India. Email: kchakraborty28@yahoo.com}

\begin{abstract}
Motivated by earlier studies \citep{tiwari,herrera}, we model
electron as a spherically symmetric charged perfect fluid
distribution of matter. The existing model is extended assuming a
matter source that is characterized by quadratic equation of state
in the context of general theory of relativity. For the suitable
choices of the parameters, our charged fluid models almost satisfy
the physical properties of electron.
\end{abstract}


\section{Introduction}

Attempts to give a classical model of charged particles, in
particular, electrons, have been going on since the time of Lorentz
\citep{lorentz}. The model proposed by Lorentz  in which electrons
have only `electromagnetic mass' and no `true' or `mechanical mass'
is commonly known as the Electromagnetic Mass Model. He assumed
electron to be an extended object consisting of pure charge and no
matter. This model was proved to be unstable but later on with the
arrival of special relativity, the model was subsequently improved.
The model was motivated from the hypothesis that gravitational mass
and other physical quantities emerge from the electromagnetic field
alone. In the past, this model has been investigated in the
classical, quantum and relativistic regimes, and we are here
re-analyzing the same model in the last scheme. The earliest
developments were made by Lorentz \citep{lorentz1} by describing the
properties of particles by their interaction with the ether.
Further, Thomson \citep{thomson} found that the kinetic energy of
the charged sphere increases by its motion through a medium of
finite specific inductive capacity, and he concluded that this
increase in kinetic energy results in the increase of mass of the
charged sphere, a phenomenon later on termed as the `electromagnetic
mass'. Later on, the improvement was made by Mie \citep{mie} through
constructing a model of electrons having their origin from the
electromagnetic fields alone. The drawbacks of Mie's model were
later removed by the Einstein's general theory of relativity, with
the inclusion of gravitational mass in the model to stabilize the
electron \citep{einstein} (see also \citep{herrera,Rahaman} for the
history of the electromagnetic mass model).

Tiwari and co-workers \citep{tiwari} gave an extensive model of
electron in the context of general relativity. They assumed electron
as charged sphere of perfect fluid and obtained the pressure inside
the charged sphere negative. It vanishes at $r = a,$ where $a$ is
the radius of the sphere. This result was later explained by Gron
\citep{gron} in the context of vacuum polarization. Tiwari
\textit{et al} showed that in their model the gravitational mass,
pressure and mass density of electron vanishes when charge is zero
\citep{tiwari1} . In a pioneering work, Bonnor and Cooperstock
\citep{bonnor} discussed the model of electron as a static charged
sphere that obeys Einstein-Maxwell theory of relativity and found
that it must contain some negative rest mass density. Moreover, they
found that the electron has negative active gravitational mass
within the sphere of radius $a = 10^{-16}$ cm. Herrera and  Varela
\citep{herrera} proposed an electromagnetic mass model satisfying
the pure charge condition, $p+ \rho = 0$, $p$ and $\rho$ being the
pressure and mass density respectively. They obtained energy density
to be negative for the radius $10^{-16}$ cm which is experimentally
verified upper limit of the radius of electron. Some authors have
also discussed the Electromagnetic mass models in different context
\citep{radin,radin1,ray06,ray08}.

Recently a new model has been proposed a quadratic EoS to describe
homogeneous and inhomogeneous cosmological models
\citep{nojiri,anand,capo}. This quadratic EoS,  $ p = p_0 +
\alpha\rho + \beta \rho^2 $, where $p_0$, $\alpha$ and $\beta $ are
parameters, is nothing but the Taylor expansion of arbitrary
barotropic EoS, $ p(\rho)$. According to \citep{anand}, the
quadratic EOS may describe dark energy or unified dark matter. In
the string theory, the gravity is a truly higher dimensional
interaction, which becomes effectively 4D at low enough energies. In
brane world models, inspired by string theory, the physical fields
in our four dimensional Universe are confined to the three brane,
while gravity can access the extra dimension. In brane world
scenario, the gravity on the brane can be described by the modified
4-dimensional Einstein's equations which contain (i) $ S_\mu^\nu$ ,
which is quadratic in the stress energy tensor of matter confined on
the brane (ii) the trace less tensor  $ E_\mu^\nu$, originating from
the 5D Weyl tensor. Although , one neglects the 5D Weyl tensor
contributions (under reasonable assumption) in these equations, in
spite of that the quadratic term of energy density appears in the
4-dimensional effective energy momentum tensor and should play a
significant role of different characteristics of the models. So, it
is not unnatural to choose quadratic form of the EoS. And this
quadratic form of the EoS is obviously the specific form of the
arbitrary barotropic EoS, $ p(\rho)$.

To search the matter source that characterized the electromagnetic
mass remains an elusive goal for theoretical physicists.  As
discussed earlier, motivated by the result in the brane world
models, we try to explore the possibility that the electromagnetic
mass be supported by quadratic EoS. We have found some specific
solutions of Einstein-Maxwell equations describing Electromagnetic
Mass for a matter source that is characterized by quadratic EoS. As
we are interested to assume the EoS is quadratic in the energy
density, we take the following form of EoS \citep{anand}, $p =
\alpha\rho + \beta \rho^2 $ where, $ \alpha, \beta $ are parameters.
At first we take these parameters as arbitrary, but restrictions on
these parameters may be specified later.

The paper is organized as follows: In Section 2, we provide the
Einstein-Maxwell field equations combined with the electromagnetic
tensor field and an isotropic fluid. The solutions with some
particular cases are given in section 3.

\section{Basic equations for constructing electromagnetic mass
model}

The static spherically symmetric space-time is given by the
line-element (in geometrized units with $G=1=c$)
\begin{equation}
ds^{2}=e^{\nu(r)}dt^{2}-e^{\lambda(r)}dr^{2}-r^{2}(d\theta^{2}+\sin^{2}\theta
d\phi^{2}), \label{Eq3}%
\end{equation}
where the arbitrary functions $\nu(r)$ and $\lambda(r)$ depend on
the radial coordinate $r$ only. As the electromagnetic mass of the
electron consists of its density and charge, the corresponding
stress-energy tensor should contain both parts accordingly, hence we
take
\begin{eqnarray}
T_{\mu\nu}^{(mass)}&=&(\rho+p)u_\mu u_\nu-pg_{\mu\nu},\\
T_{\mu\nu}^{(charge)}&=&\frac{1}{4\pi}(-F_{\mu\upsilon}F^\upsilon_\nu+\frac{1}{4}g_{\mu\nu}F_{\upsilon\chi}F^{\upsilon\chi}),\\
T_{\mu\nu}^{(total)}&=&T_{\mu\nu}^{(mass)}+T_{\mu\nu}^{(charge)}.
\end{eqnarray}
For the metric (1), the Einstein$-$Maxwell field equations are
\citep{tiwari}
\begin{equation}
e^{-\lambda}\Big[ \frac{\lambda^{\prime}}{r}-\frac{1}{r^{2}}\Big]
+\frac{1}{r^{2}}=8\pi\rho+E^{2},
\end{equation}
\begin{equation} e^{-\lambda}\Big[
\frac{1}{r^{2}}+\frac{\nu^{\prime}}{r}\Big] -\frac {1}{r^{2}}=8\pi
p-E^{2},
\end{equation}
\begin{equation}
 \frac{1}{2}e^{-\lambda}\Big[
\frac{1}{2}(\nu^{\prime})^{2}+\nu^{\prime
\prime}-\frac{1}{2}\lambda^{\prime}\nu^{\prime}+\frac{1}{r}({\nu^{\prime
}-\lambda^{\prime}})\Big]
 =8\pi p +E^{2}.
\end{equation}
Here $p(r)$, $\rho(r)$ and $E(r)$ represent the fluid pressure,
energy density and electric field, respectively, for a charged fluid
sphere. In our model, we use the quadratic equation of state
\begin{equation}
p (r)= \alpha\rho(r) + \beta [\rho(r)]^2.
\end{equation}
The electric field is expressed as
\begin{equation}
(r^{2}E)^{\prime}=4\pi r^{2}\sigma e^{\frac{\lambda}{2}}.
\end{equation}
Here $\sigma(r)$ is the charge density on the sphere. Eq. (9) gives
the following form for the electric field
\begin{equation}
E(r)=\frac{1}{r^{2}}\int_{0}^{r}4\pi r^{\prime2}\sigma e^{\frac{\lambda}{2}}%
dr^\prime=\frac{q(r)}{r^{2}},
\end{equation}
with $q(r)$ is the total charge of the sphere under consideration.
Also, the hydrodynamical equilibrium condition
($T^{\mu\nu}_{;\nu}=0$) is given by
\begin{equation} \frac{dp}{dr}
+ ( \rho + p ) \frac{ \nu^\prime }{2} = \frac{1}{ 8 \pi
r^4}\frac{dq^2}{dr}.
\end{equation}
The above expression represents the Tolman-Oppenheimer-Volkoff
equation for a charged sphere. In the next section, we are going to
solve the equations (5)$-$(10), to get the metric coefficients.
These solutions describe different type of charged fluid spheres for
given \textit{effective} densities. For a sphere of radius $r_0$,
the pressure must vanish at $r_0$ and consequently the pressure
gradient must be decreasing function of $r$. Since $\rho+p=0$ (due
to the fact $\nu+\lambda=0$), it yields that density to vanish at
the surface too. To avoid this inconsistency, it is necessary that
the sphere must be charged.

\section{Solutions of Einstein-Maxwell field equations}

One can note that  the term $  \sigma e^{\frac{\lambda}{2}} $
occurring inside the integral sign in equation (10), is equivalent
to the volume charge density. We will discuss two models assuming
the volume charge density being polynomial function of $r$. Hence we
use the condition
\begin{equation} \sigma e^{\frac{\mu}{2}} = \sigma_0 r^s, \end{equation}
where $s$ is arbitrary constant and the constant $ \sigma_0 $ is the
charge density at $r = 0$, the center of the charged matter
\citep{tiwari1,Rahaman}.

\subsection{Electron model with constant effective radial pressure}

In this specialization, we assume the effective radial pressure is
constant as
\begin{equation}
p_r^{eff}=8\pi p-E^{2}=p_o,
\end{equation}
where $p_0$ is an arbitrary constant. Here, we assume effective
radial pressure is minimum at the center, $r=0$ and equal to
constant, say $p_0$. Equation (9) together with (12) yields
\begin{equation}
E^2(r)  =  \frac{16\pi^2\sigma_o^2}{(s+3)^2} r^{2s+ 2},
\end{equation}
while Eqs. (10) and (14) give
\begin{equation}
q^2(r)  =  \frac{16\pi^2\sigma_o^2}{(s+3)^2} r^{2s+ 6}.
\end{equation}
Using equations (8), (13) and (14), one gets the following
expressions of pressure and energy density as
\begin{eqnarray}
p  &=& \frac{p_o}{8\pi}+ \frac{2\pi\sigma_o^2}{(s+3)^2} r^{2s+ 2},\\
\rho  &=& \frac{ -\alpha +
\sqrt{\alpha^2-4\beta\Big[\frac{p_o}{8\pi}+
\frac{2\pi\sigma_o^2}{(s+3)^2} r^{2s+ 2}\Big]} }{2\beta}.
\end{eqnarray}
Using the field equations and the above obtained parameters, we get
the first metric coefficient as
\begin{eqnarray}
e^\nu  &=&  \Big[(\alpha +1)\Big(\frac{ -\alpha +
\sqrt{\alpha^2-4\beta\Big[\frac{p_0}{8\pi}+
\frac{2\pi\sigma_0^2}{(s+3)^2} r^{2s+ 2}\Big]} }{2\beta}\Big)\nonumber\\
&& + \Big(\frac{ -\alpha +
\sqrt{\alpha^2-4\beta\Big[\frac{p_0}{8\pi}+
\frac{2\pi\sigma_0^2}{(s+3)^2} r^{2s+ 2}\Big]}
}{2\sqrt{\beta}}\Big)^2\Big]^{\frac{4}{(s+1)}}
\nonumber\\
&&\times \Big[\frac{\alpha +1 +\Big(\frac{ -\alpha +
\sqrt{\alpha^2-4\beta\Big[\frac{p_0}{8\pi}+
\frac{2\pi\sigma_0^2}{(s+3)^2} r^{2s+ 2}\Big]} }{2\beta}\Big) \beta
}{\Big(\frac{ -\alpha + \sqrt{\alpha^2-4\beta\Big[\frac{p_0}{8\pi}+
\frac{2\pi\sigma_0^2}{(s+3)^2} r^{2s+ 2}\Big]}
}{2\beta}\Big)}\Big]^{\frac{4}{(s+1)(\alpha+1)}}.\nonumber\\
\end{eqnarray}
Similarly, from Eq. (6) we obtain the second metric coefficient as
\begin{eqnarray}
e^{-\lambda} &=& 1-\frac{2M(r)}{r} =
1-\frac{16\pi^2\sigma_0^2}{(s+3)^2(2s+ 5)} r^{2s+ 4} \nonumber\\
&&- \frac{8 \pi}{r} \int \Big [ \frac{ -\alpha +
\sqrt{\alpha^2-4\beta\Big[\frac{p_0}{8\pi}+
\frac{2\pi\sigma_0^2}{(s+3)^2} r^{2s+ 2}\Big]} }{2\beta}\Big] r^2
dr.\nonumber\\
\end{eqnarray}
where $M(r)\equiv\int\limits_{0}^r 4\pi r^2
T^0_0dr=4\pi\int\limits_0^r r^2(\rho+E^2/8\pi)dr$, is the effective
gravitational mass. One can find the exact analytical forms of (19)
for different values of the parameter $s$.

\subsubsection{$s=0$}
\begin{eqnarray}
e^{-\lambda}&=&1-\frac{16\pi^2\sigma_o^2}{45}r^4+\frac{4\pi\alpha}{3\beta}r^2-\Big[\frac{-128\pi^3
\sigma_0^2}{9\beta}\Big]^{\frac{1}{2}}\nonumber\\
&&\times\Big[
\frac{(A^2+r^2)^{3/2}}{4}-\frac{A^2\sqrt{A^2+r^2}}{8}-\frac{A^4}{8r}\nonumber\\
&&\times\ln(r+\sqrt{A^2+r^2})\Big].
\end{eqnarray}
Here
\begin{equation}
A^2 = \frac{9(2\pi\alpha^2-p_0\beta)}{-16\pi^2\sigma_0^2\beta}.
\end{equation}
\subsubsection{$s=\frac{1}{2}$}
\begin{eqnarray}
e^{-\lambda}&=&1-\frac{32\pi^2\sigma_o^2}{147}r^5+\frac{4\pi\alpha}{3\beta}r^2-\frac{8\pi\alpha}{9\beta}
\nonumber\\
&&\frac{(\eta^\prime+r^3)^{3/2}}{r}\Big(
\frac{32\pi\beta\sigma_o^2}{49\alpha} \Big)^2.
\end{eqnarray}
Here
\begin{equation}
\eta^\prime=\frac{1+\frac{p_o\beta}{2\pi\alpha}}{\frac{32\beta\pi\sigma_o^2}{49\alpha}}.
\end{equation}
\subsubsection{$s=2$}
\begin{eqnarray}
e^{-\lambda}&=&1-\frac{16\pi^2\sigma_o^2}{225}r^8+\frac{4\pi\alpha}{3\beta}r^2-\frac{4\pi\alpha}{3\beta}\Big(
\frac{8\pi\beta\sigma_o^2}{25\alpha} \Big)^2\nonumber\\
&&\times\Big[ \frac{r^2\sqrt{r^6+k^2}}{2} +
\frac{k^2}{2r}\text{ln}(r^3+\sqrt{r^6+k^2})\Big],
\end{eqnarray}
where
\begin{equation}
k^2=\frac{1+\frac{p_o\beta}{2\pi\alpha}}{\frac{8\pi\beta\sigma_o^2}{25\alpha}}.
\end{equation}
\subsubsection{$s=-\frac{1}{2}$}
\begin{eqnarray}
e^{-\lambda}&=&1-\frac{16\pi^2\sigma_o^2}{25}r^3+\frac{4\pi\alpha}{3\beta}r^2\nonumber\\
&&-\frac{4\pi\alpha}{\beta}\frac{2(15a^2r^2-12abr+8b^2)}{105a^3r}(ar+b)^{3/2}.\nonumber\\
\end{eqnarray}
where
\begin{equation}
a=\frac{32\pi\beta\sigma_o^2}{25\alpha},\ \
b=1+\frac{p_o\beta}{2\pi\alpha}.
\end{equation}
\subsubsection{$s=-2$}
\begin{equation}
e^{-\lambda}=1-16\pi^2\sigma_o^2+\frac{4\pi\alpha}{3\beta}r^2+\frac{4\pi\alpha}{3\beta
a}\frac{(ar^2+b)^{3/2}}{r},
\end{equation}
where
\begin{equation}
a=1+\frac{p_o\beta}{2\pi\alpha},\ \
b=\frac{8\pi\beta\sigma_o^2}{\alpha}.
\end{equation}

\subsection{Electron model with constant effective energy density}

In this specialization, we assume the effective energy density is
constant as
\begin{equation}
\rho_{eff} = 8\pi\rho+E^{2} =  \rho_o
\end{equation}
where $\rho_o$ is an arbitrary constant. We assume here the
effective energy density has just opposite behavior to the previous
case i.e. it is maximum at the center, say $\rho_0$. In this case
the forms of $E$ and $q$ are same as (14) and (15). Following the
procedure in the previous section, the other parameters can be found
as
\begin{eqnarray}
\rho &=& \frac{\rho_0}{8\pi} - \frac{2\pi\sigma_0^2}{(s+3)^2}
r^{2s+2},\\
p &=& \alpha \Big[\frac{\rho_0}{8\pi} -
\frac{2\pi\sigma_0^2}{(s+3)^2} r^{2s+ 2} \Big] \nonumber\\
&&+ \beta \Big[\frac{\rho_0}{8\pi}- \frac{2\pi\sigma_0^2}{(s+3)^2}
r^{2s+
2}\Big]^2,\\
e^{-\lambda} &=& 1-\frac{\rho_0 r^2}{3},\\
\nu&=&\int re^\lambda(8\pi p-E^2+\frac{1}{r^2})dr-\int\frac{dr}{r},
\end{eqnarray}
which gives
\begin{eqnarray}
\nu&=&\alpha\rho_o\int\frac{1}{1-\frac{\rho_or^2}{3}}dr+\frac{\beta\rho_o^2}{8\pi}\int\frac{1}{1-\frac{\rho_or^2}{3}}dr\nonumber\\
&&-\frac{8\pi\sigma_o^2(2\pi\alpha+\frac{\beta\rho_o}{2})}{(s+3)^2}\int\frac{r^{2s+3}}{(1-\frac{\rho_or^2}{3})}dr\nonumber\\
&&+\frac{32\pi^3\sigma_o^4\beta}{(s+3)^2}\int\frac{r^{4s+5}}{1-\frac{\rho_o^2r^2}{3}}dr-\frac{16\pi^2\sigma_o^2}{(s+3)^2}\int\frac{r^{2s+3}}{1-\frac{\rho_o^2r^2}{3}}dr\nonumber\\
&&-\int\frac{dr}{r(1-\frac{\rho_o^2r^2}{3})}-\int\frac{dr}{r}.\nonumber\\
\end{eqnarray}
On simplification, we obtain
\begin{eqnarray}
\nu&=&(\alpha\rho_o+\frac{\beta\rho_o^2}{8\pi})\sqrt{\frac{3}{\rho_o}}\text{Tanh}^{-1}(r\sqrt{\frac{\rho_o}{3}})
\nonumber\\
&&-\frac{8\pi\sigma_o^2(2\pi\alpha+\frac{\beta\rho_o}{2})}{(s+3)^2}\frac{r^{4+2s}}{2(2+s)}{}_{2}F_{1}(2+s,1,3+s,\frac{r^2\rho_o}{3})\nonumber\\
&&+\frac{32\pi^3\sigma_o^4\beta}{(s+3)^2}\frac{3r^{6+4s}}{18+12s}{}_{2}F_{1}(3+2s,1,4+2s,\frac{r^2\rho_o^2}{3})
\nonumber\\
&&-\frac{16\pi^2\sigma_o^2}{(s+3)^2}\frac{r^{4+2s}}{2(2+s)}{}_{2}F_{1}(2+s,1,3+s,\frac{r^2\rho_o^2}{3})\nonumber\\
&&-2\text{ln}r+\frac{1}{2}\text{ln}(r^2\rho_o^2-3).
\end{eqnarray}
Here we are interested in the the case for $s=0$, which yields
\begin{eqnarray}
\nu&=&(\alpha\rho_o+\frac{\beta\rho_o^2}{8\pi})[\sqrt{\frac{3}{\rho_o}}\text{Tanh}^{-1}(r\sqrt{\frac{\rho_o}{3}})]
\nonumber\\
&&-\frac{8\pi\sigma_o^2(2\pi\alpha+\frac{\beta\rho_o}{2})}{9}\frac{r^{4}}{4}{}_{2}F_{1}(2,1,3,\frac{r^2\rho_o}{3})\nonumber\\
&&+\frac{32\pi^3\sigma_o^4\beta}{9}\frac{3r^{6}}{18}{}_{2}F_{1}(3,1,4,\frac{r^2\rho_o^2}{3})
\nonumber\\
&&-\frac{16\pi^2\sigma_o^2}{9}\frac{r^{4}}{4}{}_{2}F_{1}(2,1,3,\frac{r^2\rho_o^2}{3})
\nonumber\\
&&-2\text{ln}r+\frac{1}{2}\text{ln}(r^2\rho_o^2-3).
\end{eqnarray}

\section{Discussions}

We have provided two new toy electromagnetic mass models which
suggest that mass of electron has the ultimate origin from the
electromagnetic field alone. For this, we have considered a static
spherically symmetric charged perfect fluid with the quadratic  EoS
$ p = \alpha\rho + \beta \rho^2 $. In the first model, the total
effective pressure is taken to be a combination of the barotropic
pressure and the one generated due to the electric field. While in
the second model, the effective energy density is taken to be sum of
barotropic energy density and due to the electric field. The
parameter in the EoS, namely $\alpha$ is related to sound speed for
the fluid as follows: The squared of sound velocity, $v_s^2 =
\partial p/\partial\rho = \alpha + 2 \beta \rho $, is always
positive irrespective of matter density. In literature, more
generally employed EoS parameter $\omega$ is: $ \omega=p/\rho =
\alpha +  \beta \rho \alpha  $. Hence, $\alpha$ is related to sound
speed and EoS parameter as $ \alpha = 2\omega-v_s^2 $.

According to previous models \citep{herrera,bonnor}, an electron
(experimentally obtained  radius $ r \sim 10^{-16} $ cm, the
inertial mass $ m  \sim 10^{-56} $cm, charge $ Q \sim 10^{-34} $cm
in relativistic units)  is modeled as a charge fluid sphere obeying
Einstein-Maxwell theory. So, the interior metric should be matched
with the exterior Reissner-Nordstr\"{o}m metric. If $M$ be the
effective gravitational mass within the fluid sphere, then matching
at the junction interface implies $ M = m - \frac{Q^2}{2a} $.
Putting the above values of inertial mass, charge and radius of the
electron, one gets the value of the effective gravitational mass
within the fluid sphere is $ \sim - 10^{-52}$cm.

Since, we have studied two models assuming the volume charge density
being polynomial function of $r$, so for both the models the
electric field and total charge of the sphere under consideration
are the same. If we assume the charge density $\sigma_0 $ at $ r = 0
$, the center of the charged matter is $ \sigma_0   \sim 10^{12} $,
then charge of the fluid is nearly equal to $ q  \sim 10^{-34} $cm
in relativistic units (see the fig 1) which is equivalent to the
charge of electron. In a similar way, if one assumes the suitable
values of the parameters, then all physical parameters like,
pressure, density and effective gravitational mass of the charged
fluid that represents the electron can be obtained (see figs. 3, 4
and 5).

In model 1, one can note that for the following values of the
parameters: $\sigma_0\sim 10^{12}$, $s=0$, $ \alpha = -10^{-4}$,
$\beta = -0.1$, the term $\frac{4\pi\alpha}{3\beta}r^2$ is the
dominating term in equation (20). In this case the effective
gravitational mass $ \sim - 10^{-52} $ (see the figure 5). Note
that this result of negative mass is consistent with that of [12],
if the size of electron is taken to be less then $10^{-16}$cm.
This negative mass and gravitational repulsion is a signature of
the strain of the vacuum due to vacuum polarization [8]. For model
2, if one chooses the values of the parameters $\rho_0 = -
10^{-4}$ and $s=0$ , then the effective gravitational mass $ \sim
- 10^{-52} $ (see the figure 8). In figs. (4) and (7), we see that
the energy density $\rho$ has decreasing behavior against the
radial parameter and it takes negative values. This behavior
arises since $\rho=-p$ for $p>0$ (vacuum fluid). Thus figures (3)
and (6) manifest the rising behavior of pressure.

In the present work, we have considered charged fluid  of radius
$\sim 10^{-16} $ ( i.e. electron ) obeying quadratic
 equation of state. We hope other people would be motivated by our approach and in
future will try to extrapolate the present investigation to the
astrophysical bodies, specially quark or strange stars.

\subsubsection*{Acknowledgments}
FR is thankful to PURSE for providing financial support.

\pagebreak

\begin{figure}[htbp]
    \centering
        \includegraphics[scale=.35]{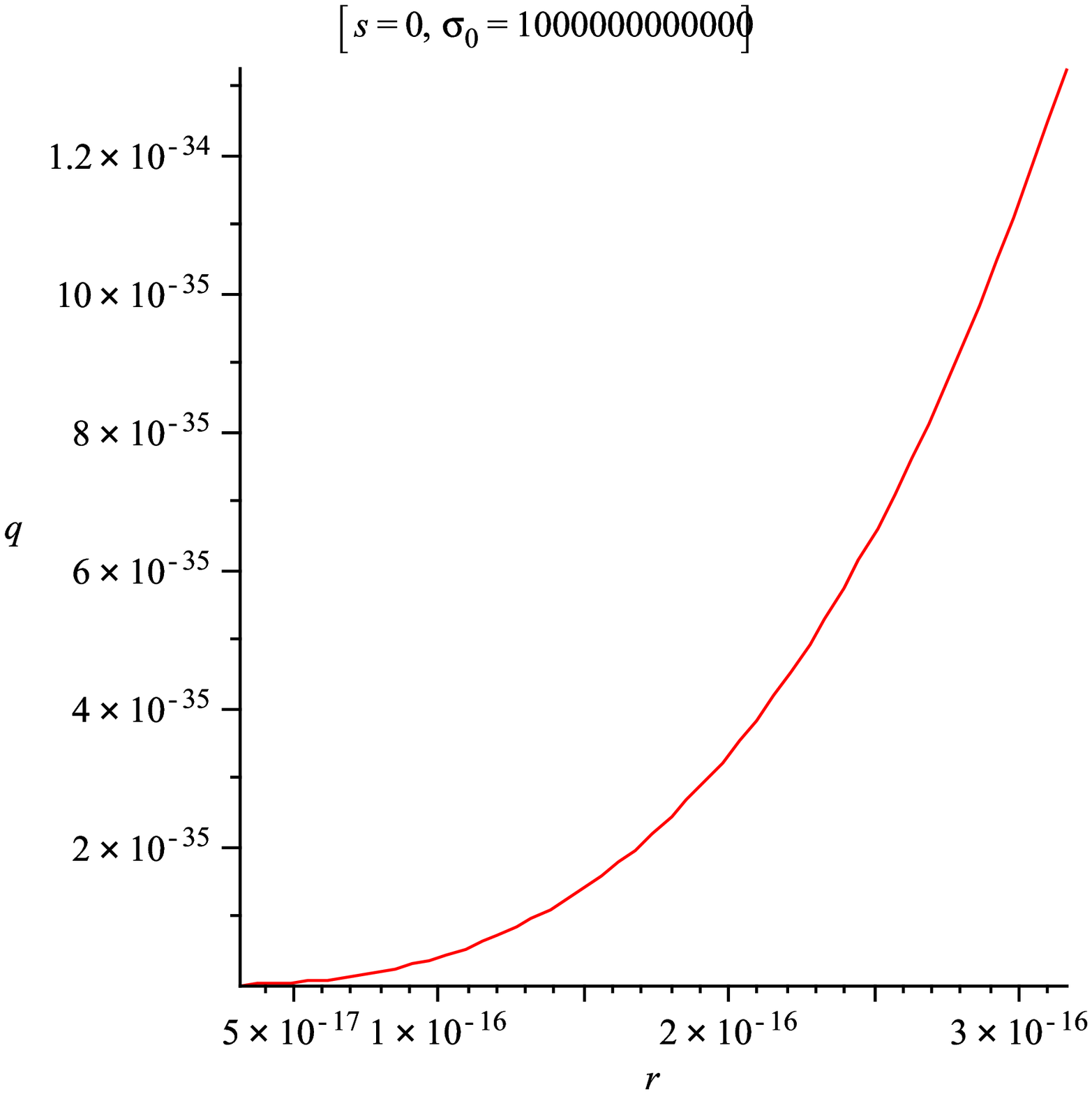}
        \caption{ The diagram of the Electric charge $q$ with respect to
radial coordinate $r$ for  $s = 0$.}
\end{figure}

\begin{figure}[htbp]
    \centering
        \includegraphics[scale=.35]{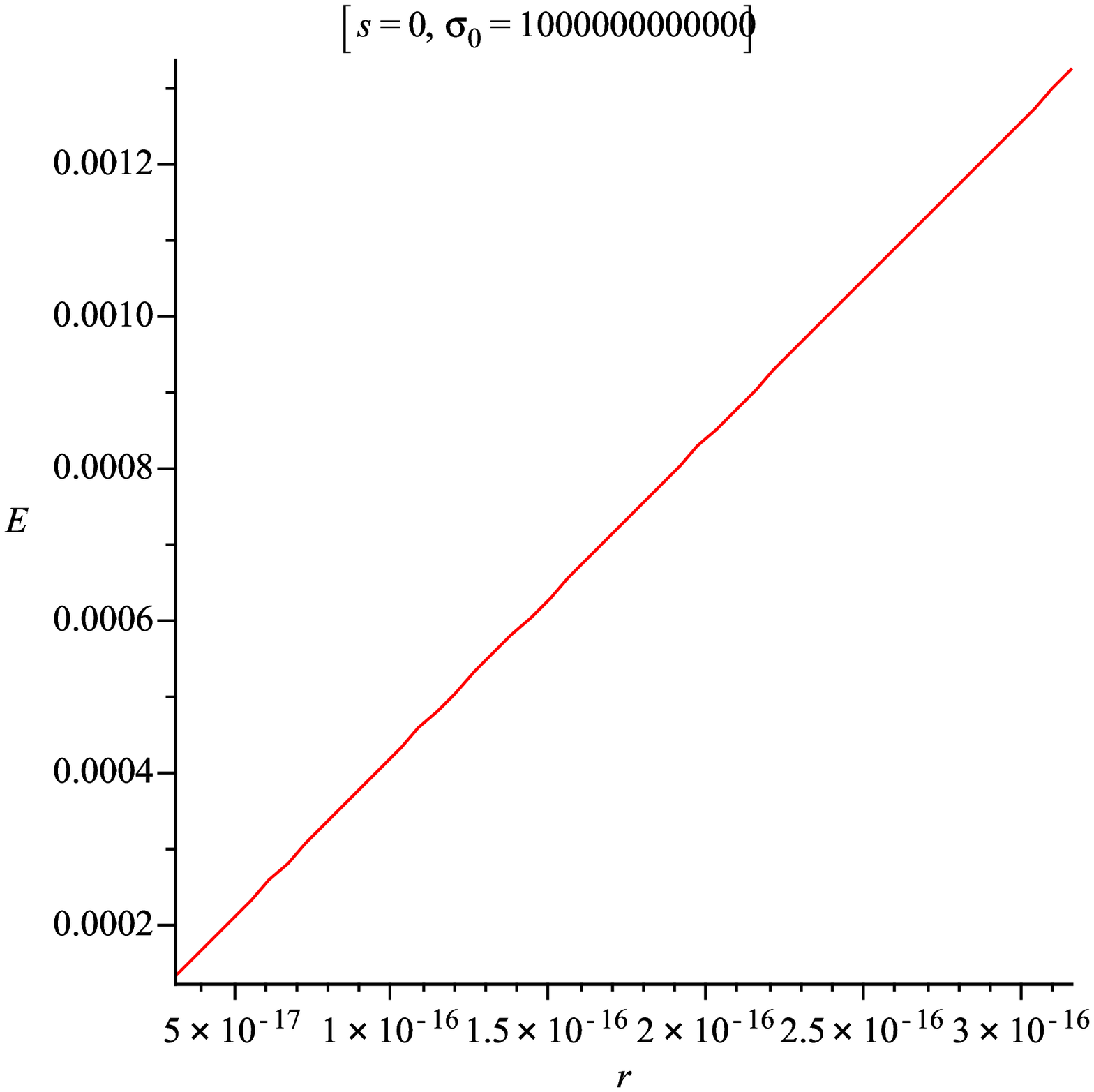}
        \caption{The diagram of the Electric field strength $E$ with
respect to radial coordinate $r$ for  $s = 0$.}
\end{figure}

\begin{figure}[htbp]
    \centering
        \includegraphics[scale=.3]{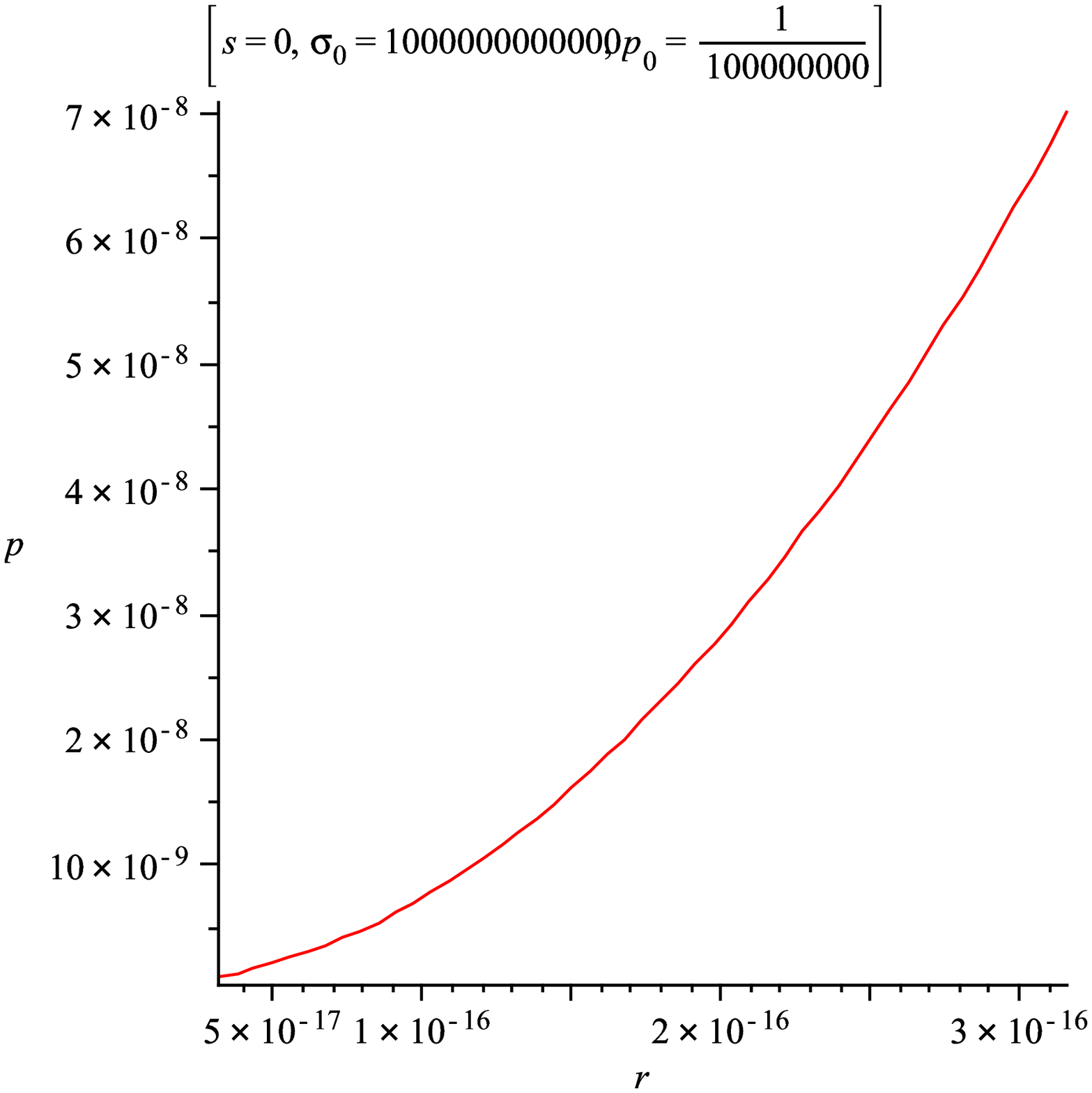}
        \caption{The diagram of the radial pressure $p$ with respect to
radial coordinate $r$ for  $s = 0$. }
\end{figure}

\begin{figure}[htbp]
    \centering
        \includegraphics[scale=.3]{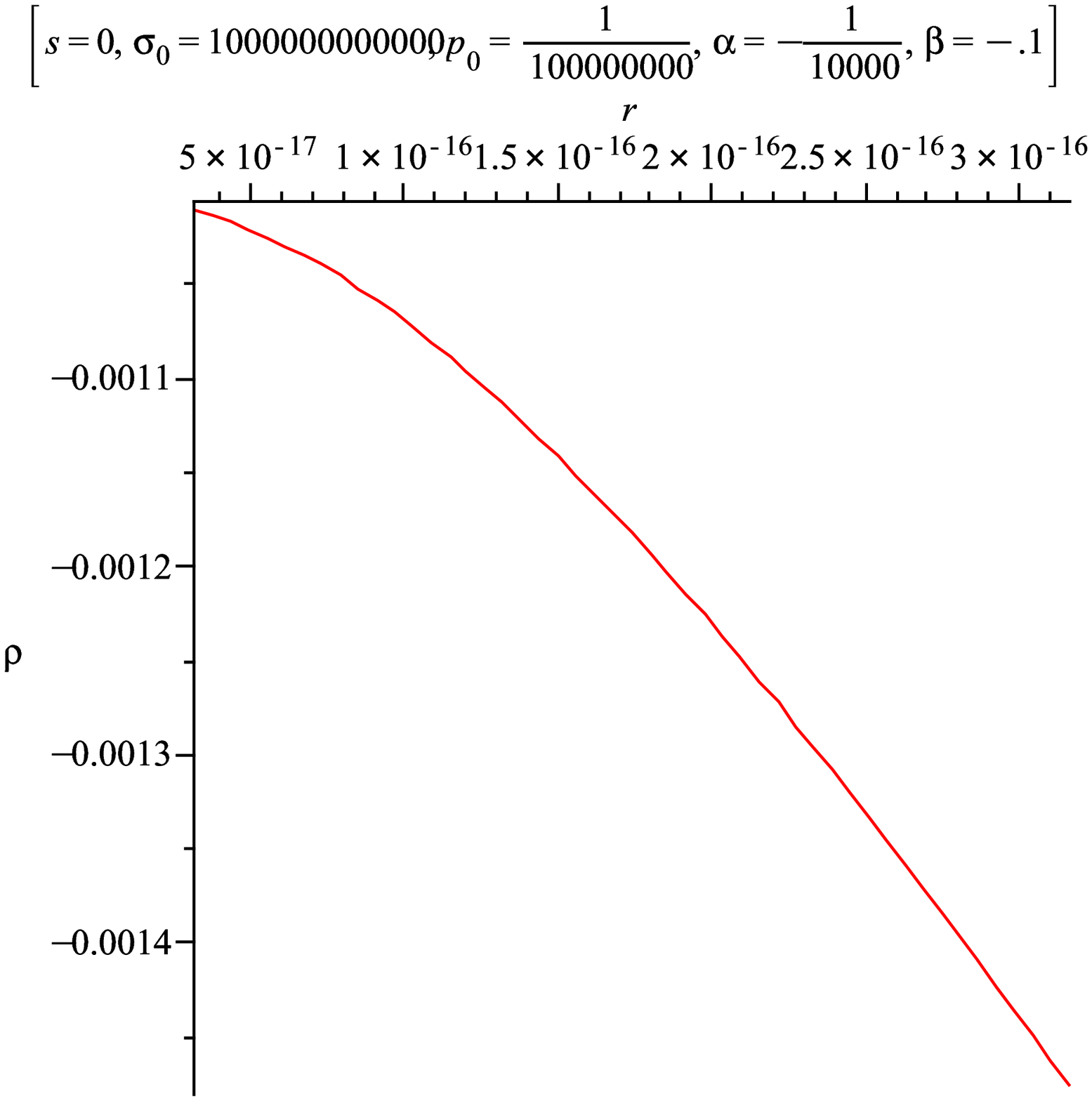}
        \caption{ The diagram of the energy density  with respect to
radial coordinate $r$ for  $s = 0$.}
\end{figure}

\begin{figure}[htbp]
    \centering
        \includegraphics[scale=.3]{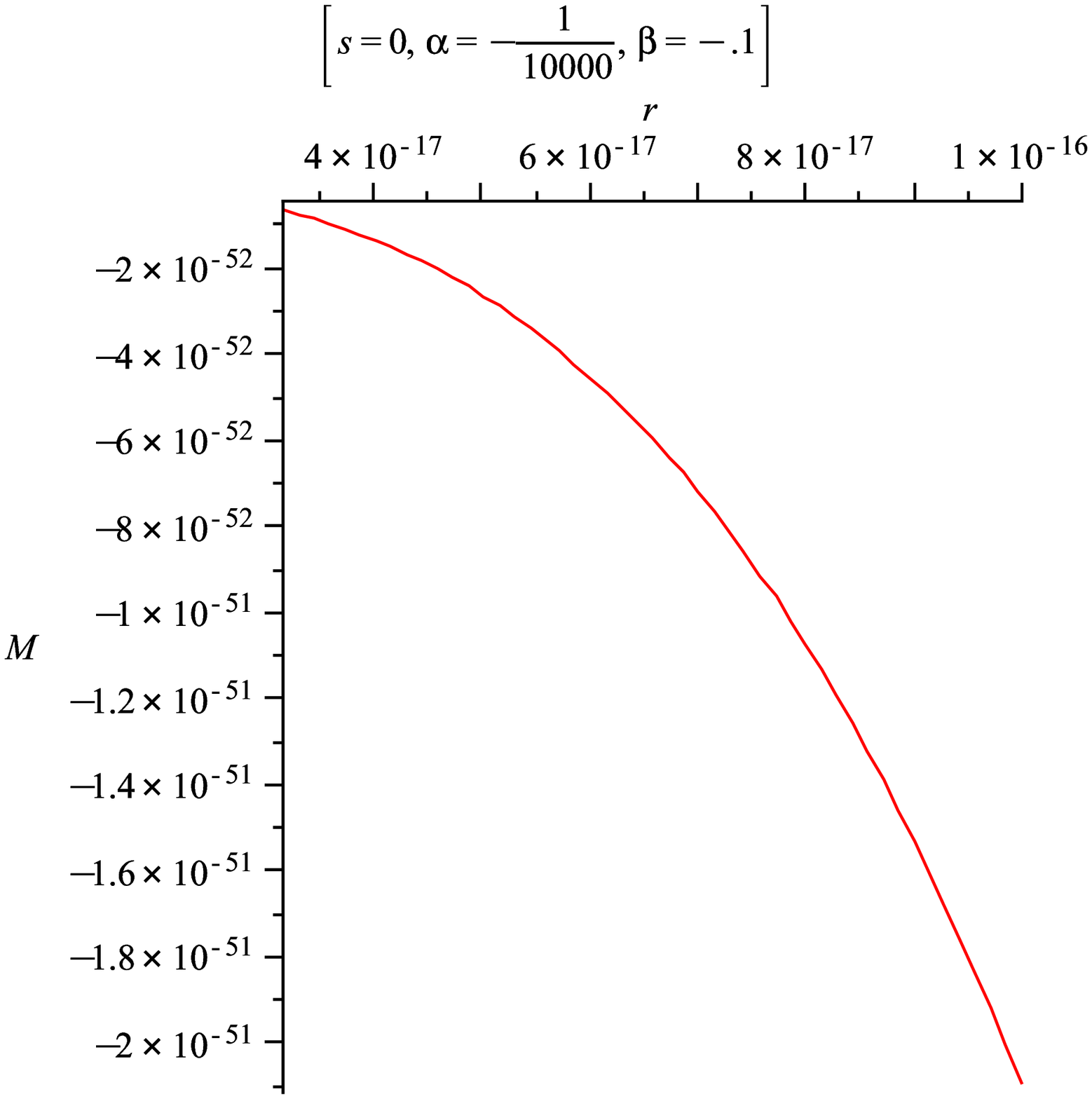}
        \caption{ The diagram of the effective gravitational mass
function $M$ with respect to radial coordinate
        $r$ for suitable values of the parameters. }
\end{figure}

\begin{figure}[htbp]
    \centering
        \includegraphics[scale=.3]{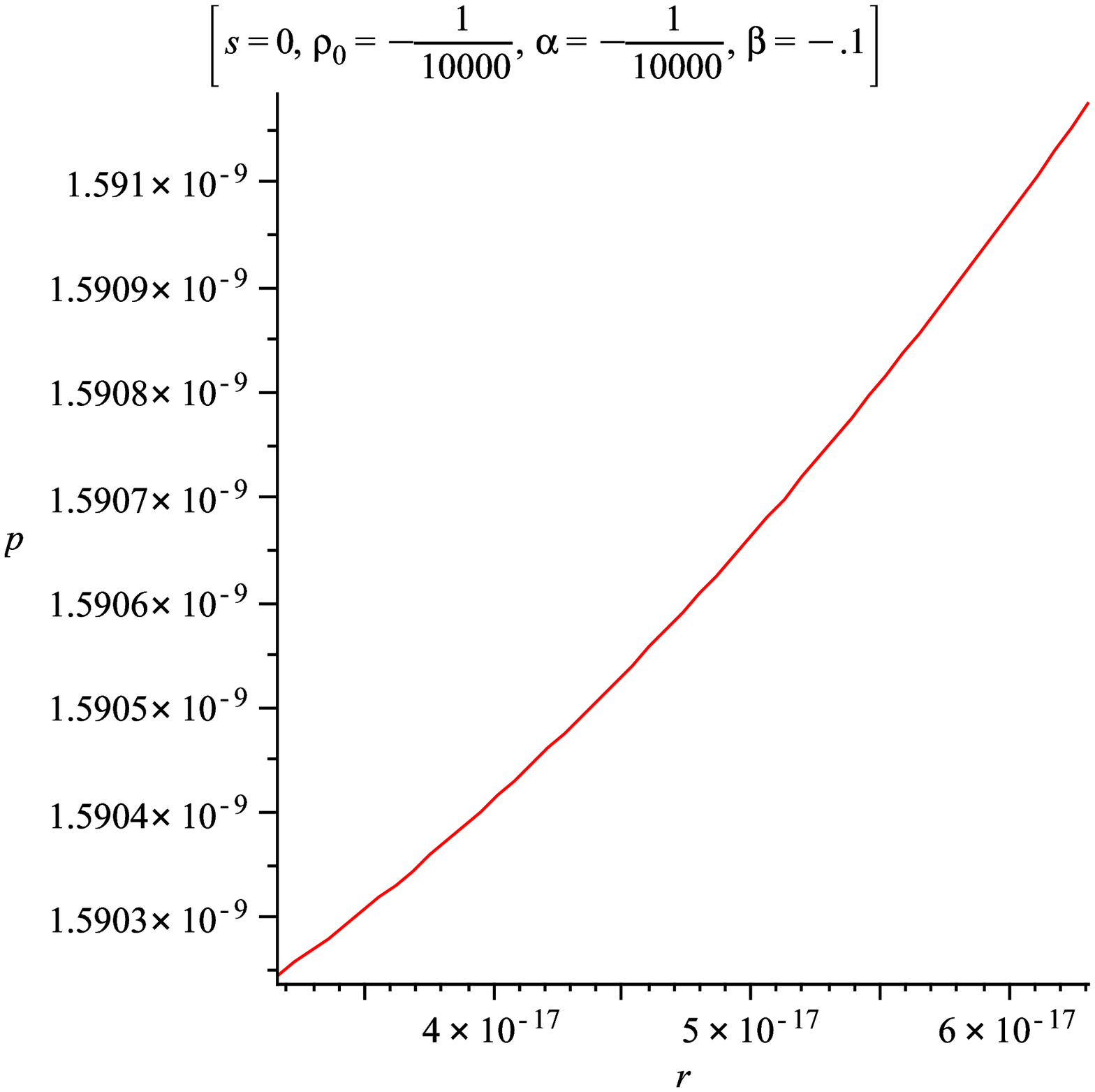}
        \caption{The diagram of the  radial pressure $p$ with respect to
radial coordinate $r$  for $ s = 0$ for model 2. }
\end{figure}

\begin{figure}[htbp]
    \centering
        \includegraphics[scale=.3]{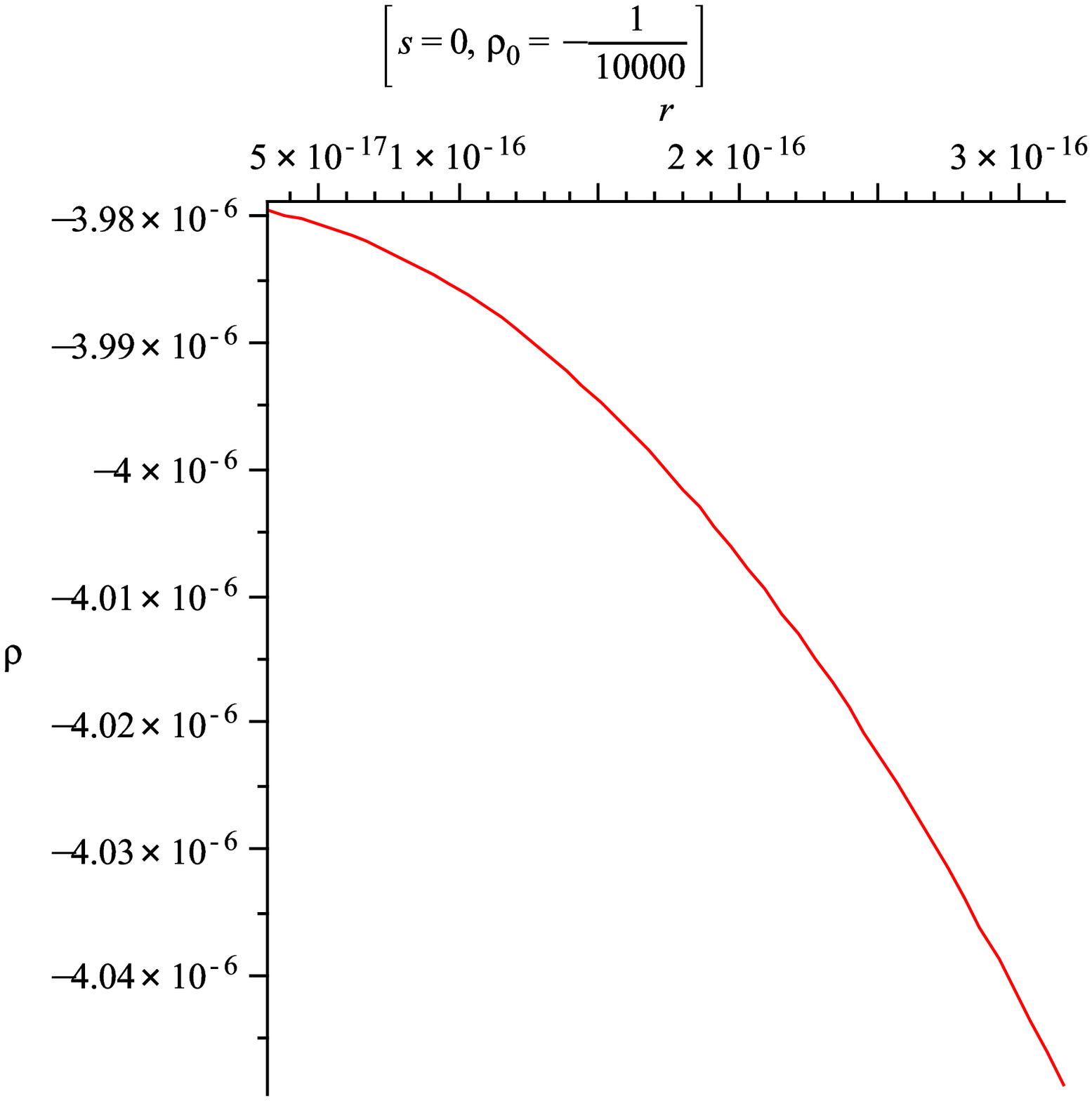}
        \caption{ The diagram of the energy density $\rho$ with respect
to radial coordinate $r$  for  $s = 0$ for model 2.}
\end{figure}

\begin{figure}[htbp]
    \centering
        \includegraphics[scale=.3]{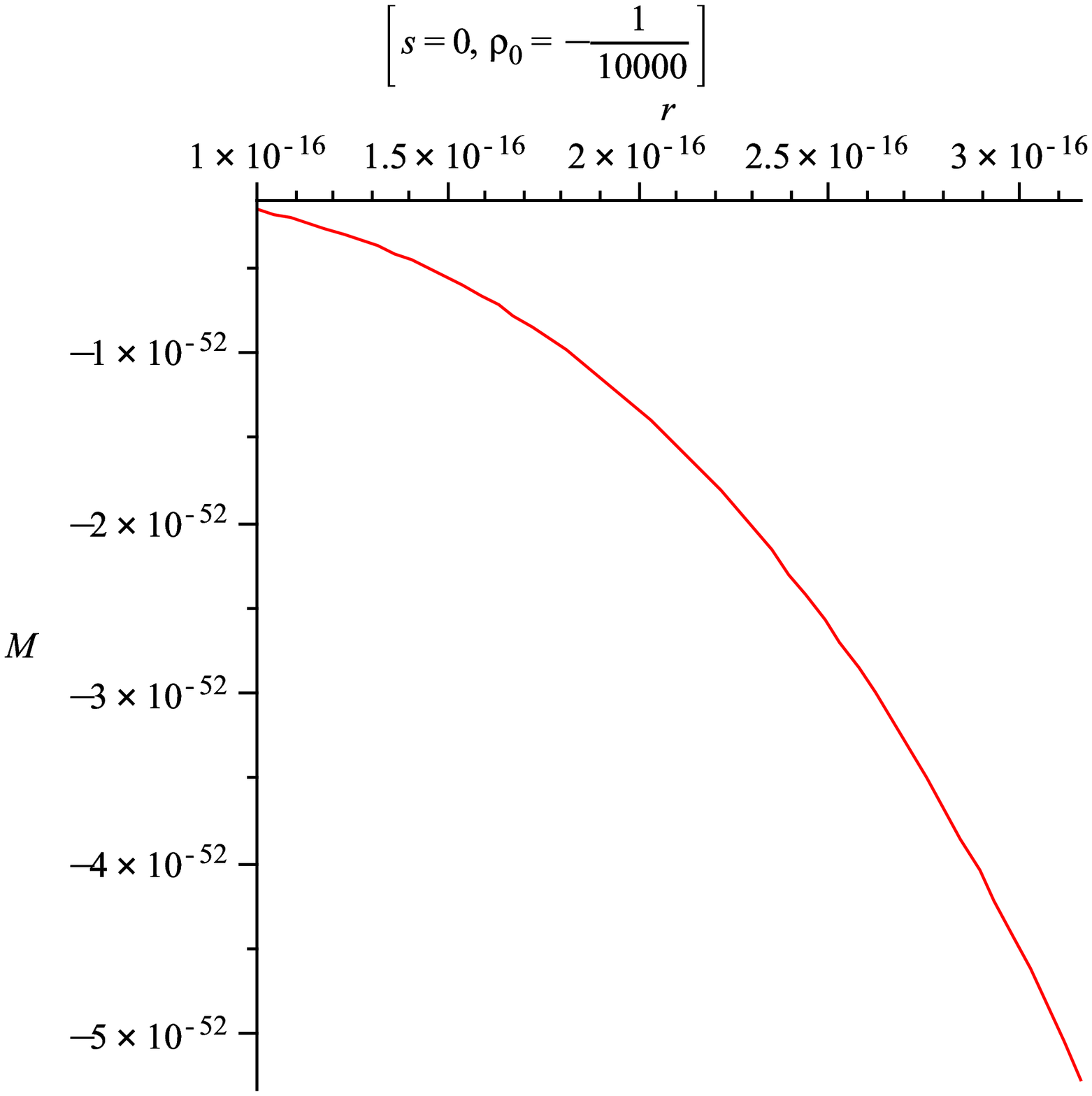}
        \caption{ The diagram of the effective gravitational mass
function $M$ with respect to radial coordinate
        $r$ for the model 2 for suitable value of the parameter, $\rho_0 = - 10^{-4}$. }
\end{figure}

\end{document}